\documentclass[twocolumn,showpacs,preprintnumbers,amsmath,amssymb]{revtex4}
\usepackage{graphicx}% Include figure files
\usepackage{dcolumn}% Align table columns on decimal point
\usepackage{bm}% bold math
\begin{document}
%\draft

\title{Elementary Electronic  Excitations  in  Graphene Nanoribbons.}

\author{L. Brey$^1$ and H. A. Fertig$^{2,3}$}
\affiliation{1. Instituto de Ciencia de Materiales de Madrid (CSIC), Cantoblanco, 28049 Madrid, Spain\\
2. Department of Physics, Indiana University, Bloomington, IN 47405\\
3. Department of Physics, Technion, Haifa 32000, Israel}

\date{\today}

\begin{abstract}
We analyze the collective mode spectrum of
graphene nanoribbons within the random phase approximation.
In the undoped case, only metallic
armchair nanoribbons support a propagating plasmon mode.
Landau damping of
this mode is shown to be suppressed through the chirality of the
single particle wavefunctions.
We argue that
undoped zigzag nanoribbons should not support plasmon excitations
because of a broad continuum of particle-hole excitations
associated with surface states, into which collective
modes may decay.  Doped nanoribbons have properties
similar to those of semiconductor nanowires, including a plasmon
mode dispersing as $q\sqrt{-\ln qW}$  and
a static dielectric response that is divergent at $q=2k_F$.

\end{abstract}
%\pacs{73.43-f,73.20-r,73.23-b}
\maketitle

\section{Introduction}
Recent improvements in the processing of graphite have made
possible the isolation of single graphene
layers  \cite{Novoselov_2004}. These carbon sheets are very
promising for microelectronic applications because they support the electric
field effect: by applying a gate voltage a high mobility two
dimensional electron or hole gas can be
created \cite{Novoselov_2005,Zhang_2005}. The graphene sheets
could be further processed into multiterminal devices with
graphene nanoribbons acting as wires connecting different
nanodevices. This possibility has motivated much recent
work on carbon-based
nanoelectronics.

The electronic properties of graphene nanoribbons depend strongly on
their size and geometry \cite{Ezawa_2006,Brey_2006a}.  This dependence
occurs in part because edges of the ribbons parallel
to different symmetry directions require
different boundary conditions.
For nanoribbons with zigzag  edges (``zigzag ribbons'') the boundary
conditions lead to a particle-like and a hole-like band with
evanescent wavefunctions confined to the surfaces \cite{Brey_2006a},
which continuously  evolve into the well-known zero energy surface
states \cite{Fujita_1996,Peres_2006} as the nanoribbon width gets
larger. In the case of  armchair  edge nanoribbons
(``armchair ribbons'') the band
structure is metallic when the width of the sample in lattice
constants units has the form 3$M$+1, with $M$ an integer, and
insulating otherwise.

With respect to the lowest energy states,  graphene nanoribbons are
one dimensional (1D) systems. However, their band structure is
rather different than that of more conventional 1D systems.
In armchair metallic nanoribbons the dispersions of the electron
and hole bands are linear in momentum, whereas in armchair
semiconductor nanoribbons  the  low energy  dispersion is quadratic.
On the other hand,  zigzag nanoribbons have a peak in the density of
states at the Fermi energy. This affects the low energy
physics and the screening properties of graphene nanoribbons. In
this work we compute, within the Random Phase Approximation
(RPA) \cite{Mahan_book}, the dynamical polarizability and the
dielectric constant of a doped graphene nanoribbon.  We calculate
the low energy charged collective excitations, and
analyze the differences between conventional semiconductor-based one dimensional
systems and graphene nanoribbons.

We now summarize our results.
Within the RPA, we find that
for most undoped armchair  nanoribbons there are no
plasmon modes associated with interband transitions.
The only exception is
metallic
nanoribbons, which support a plasmon that disperses  as $q \sqrt {-\ln q}$,
with $q$ the momentum of the excitation along the nanoribbon.
Unlike plasmons in semiconductor based 1D systems, in metallic
nanoribbons the energy of the  plasmons  does not depend on the
level of doping of the system.  The case of
undoped zigzag
nanoribbons turns out to be considerably more
complicated, and we do not
carry through a full RPA analysis as in
the armchair chase.  However, we argue that a broad particle-hole
continuum in the zigzag case implies that any poles in the
inverse dielectric function will be damped, so that
there cannot be propagating plasmon modes.  Finally,
we show that doped semiconductor
nanoribbons support low energy plasmons that disperse as $q \sqrt {-\ln
q}$ and, in the armchair case, depend on the density of carriers in the ribbon in
a standard way.

This paper is organized as follows. In Section II we
introduce the Dirac Hamiltonian used for describing the electronic
properties of graphene. In  subsections II-A and II-B we review the
non-interacting electronic structure of armchair and zigzag nanoribbons. Section
III is dedicated to defining a generalized dielectric function for
graphene nanoribbons, and to presenting results for collective
charge excitations in intrinsic (undoped) and doped graphene
nanoribbons which may be derived from it.
We conclude in Section IV with a summary of the results.

\section{Model Hamiltonian.}
Graphene is a honeycomb structure of covalently
bonded carbon atoms, which should be treated as a triangular lattice
with two basis atoms, denoted by $A$ and $B$, per unit cell. In graphene
the low energy dispersions of the electron and hole bands are linear
near the points ${\bf K}=\frac{2
\pi}{a_0}(\frac{1}{3},\frac{1}{\sqrt{3}})$ and ${\bf K}'=\frac{2
\pi}{a_0}(-\frac{1}{3},\frac{1}{\sqrt{3}})$ of the Brillouin zone.
Here $a_0$ is the triangular lattice parameter of the graphene
structure. In the ${\bf k} \cdot {\bf p}$
approximation \cite{Ando_2005,DiVincenzo_1984} the wavefunctions are
expressed in terms of envelope functions $[\psi_A({\bf
r}),\psi_B({\bf r})]$ and $[\psi_A^{\prime}({\bf
r}),\psi_B^{\prime}({\bf r})]$ for states near the ${\bf K}$ and
${\bf K}^{\prime}$ points, respectively. The envelope wavefunctions
may be combined into a four vector which satisfies the Dirac
equation $H\Psi = \varepsilon \Psi$, with
\begin{equation}
H=\gamma a_0\,
\left( \begin{array} {cccc} 0 & k_x-ik_y & 0 & 0 \\
k_x+ik_y & 0 & 0 & 0 \\
0 & 0& 0 & -k_x-ik_y \\
0 & 0 & -k_x+ik_y & 0 \
\end{array} \right) \, \, \, , \label{hamilt_kp}
\end{equation}
where $\gamma=\sqrt{3}t/2$, with $t$ the hopping amplitude between
nearest neighbor carbon atoms. Note that ${\bf k}$ denotes the
separation in reciprocal space of the wavevector from the ${\bf K}$
(${\bf K}^{\prime}$) point in the upper left (lower right) block of
the Hamiltonian.

The bulk solutions of Hamiltonian (\ref{hamilt_kp})  have energies
$\varepsilon$=$s \gamma a_0 |{\bf k}|$ with $s$=$\pm 1$ and are
degenerate in the valley index. The wavefunctions take the form $
[e ^{- i \theta _{{\bf k}}},s , 0,0 ]e ^{i {\bf k} {\bf r}}/C $ for
the $\bf K$ valley, and $\ [0,0, e ^{ i \theta _{{\bf k}}} ,s]e ^{i
{\bf k} {\bf r}}/C$ for the ${\bf K}'$ valley. Here $\theta _{\bf
k}= \arctan {k_x/k_y}$, and
 $C=\sqrt{2 L_x
L_y}$ is a normalization constant, with $L_x$ and $L_y$ the sample
dimensions.

\subsection{Armchair nanoribbon.}
In an armchair nanoribbon  the
termination at the edges consists of a line of $A$-$B$ dimers, so that the
wavefunction amplitude should vanish on both sublattices at the
extremes, $x$=0 and $x$=$W+a_0/2$, of the nanoribbon. To satisfy
this boundary condition we must admix valleys \cite{Brey_2006a}, and
the confined wavefunctions have the form,
\begin{equation}
\Psi _{n,s} =
\frac{e ^{i k_y y}}{2\sqrt{W+a_0/2} \sqrt{ L_y }}  \left ( \begin{array}{c} e^{-i \theta _{k_n, k_y} } \, e ^{i k_n x} \\
s \, e ^{i k_n x} \\ e^{-i \theta _{k_n, k_y} }  \, e ^{-i k_n x} \\
-s \, e ^{-i k_n x}
\end {array} \right ) \label{wf_AC}
\end{equation}
with energies $\varepsilon_{n,s} (k_y)$=$s a _0 \gamma \sqrt {  k _n
^2 + k _y ^2 }$. The allowed values of  $k_n$ satisfy the
condition \cite{Brey_2006a}
\begin{equation}
k_n = \frac {2\pi n }{2W+a_0} \,   + \frac{2\pi}{3a_0}
\, \, \, \, . \label{condition1}
\end{equation}
For a width of the form $W$=$(3M+1)a_0$, the allowed values of
$k_n$, $k_n = \frac{2\pi}{3a_0}\left(\frac{2M+1+n}{2M+1}\right)$,
create doubly degenerate states for
$n \ne -2M-1$, and allow a zero energy state when $k_y \rightarrow
0$. Nanoribbons of widths that are not of this form have
nondegenerate states and do not include a zero energy mode. Thus
these nanoribbons are band insulators. The allowed values of $k_n$
are independent of the momentum $k_y$, and the coupling between the
motion in the $x$ and $y$ directions only enters in the direction of
the isospin of the state, $\theta _{k_n, k_y}$.

\subsection{Zigzag nanoribbon.}
The boundary conditions for zigzag nanoribbons are vanishing of
the wavefunction on the $A$ sublattice at one edge, $x$=0,
and on the $B$ sublattice at the other, $x$=$W$. The geometry of the
zigzag nanoribbon does not mix the valleys, and solutions near the
$\bf K$ valley with wavevector $k_y$ are degenerate with solutions
near the $\bf K'$ valley with  wavevector -$k_y$.

For the $\bf K$ valley and  a given value of $k_y$ the nanoribbons
wavefunctions take the form
\begin{equation}
\Psi _{n,s,k_y} =   \left ( \! \begin{array}{c} i \, \, s \sinh (z_n
x)
\\ \sinh ((W-x)z_n) \end{array} \right ) \frac {e ^ {i k_y y}}{C}
\label{zigzagWF}
\end{equation}
where $C$ is the appropriate normalization constant. The
corresponding eigenvalues are   $\varepsilon = s \gamma a_0 \sqrt
{k_y ^2 - z_n ^2}$ with $s$=$\pm1$. The boundary conditions lead to
a transcendental equation for the allowed values of
$z_n$,
\begin{equation} \frac { k _y -z_n } { k _y
+z_n} = e ^{ -2 W z_n} \, \, \, . \label{trans}
\end{equation}
For $k_y >1/W$, equation (\ref{trans}) has solutions with real
values of $z$ which correspond to states localized at the surface.
There are two branches of such surface states, one with positive
energies, and the other with negative energies.
These energies approach zero as $k_y$
increases. In undoped graphene the Fermi
energy is located between these surface state branches.
For $k_y <0$, there are no solutions with
real $z$ and surface states are absent. The solutions
to Eq. \ref{trans}
have purely imaginary $z$
and represent confined states. For values of $k_y$ in
the range $0<k_y<1/W$, surface states
from the two edges
are strongly mixed and
are indistinguishable from confined states. In zigzag
nanoribbons the allowed values of $z_n$ and the corresponding
wavefunctions in the transverse direction $x$ depend strongly on
the value of the longitudinal momentum of the state $k_y$. This is a
consequence of the chirality implicit in  the Dirac Hamiltonian,
Eq.(\ref{hamilt_kp}).

\section{Dielectric constant.} In nanoribbons the electrons are
confined to move just in the $y$ direction, and the momentum in this
direction, $k_y$, is a good quantum number. The motion in the $x$
direction is confined and it is described by a band index $n$ and the
conduction or valence character $s$. The form of the dielectric
constant depends strongly on whether the motion in the longitudinal and
transverse directions of the wire  are coupled,
as in the case of the zigzag
nanoribbon, or decoupled, as in the armchair nanoribbon. In the latter
case the dielectric constant can be described as a tensor in the
subband indices whereas in the former case it also depends on the
momentum $k_y$ of the electronic states.  For undoped ribbons, we will analyze the
armchair case in detail, whereas for the zigzag case we will argue
that propagating plasmons are unlikely to exist.  The results
for doped graphene nanoribbons are qualitatively similar to
those of standard semiconductor nanowires
for both cases.

\subsection{Armchair nanoribbon.}
For armchair nanoribbons the motion of the carriers on the wire is
coupled with the transverse motion only through the relative phases
of the wavefunctions [cf. Eq.(\ref{wf_AC})]. In this situation the
generalized dielectric function \cite{DasSarma_1985,DasSarma_1984}
for the  nanoribbon is given by
\begin{equation}
\varepsilon _{ijmn} (q,\omega)=\delta_{im}\delta_{jn}- v_{ijmn}(q)
\Pi _{m,n} (q,\omega) \label{diel}
\end{equation}
where the function $\Pi _{mn} (q,\omega)$ is a generalized irreducible
polarizability function.  In this expression $i,j,m,n$ are composite
indices specifying the subband index and the valence/conduction band
character $s=\pm 1$.  The wavevector $q$ is along the direction in
which the electron motion is free, and $\delta_{im}$ is a Kronecker delta
function.  The explicit expression for $\Pi$ takes the form
\begin{eqnarray}
%\Pi _{ns,n's'} (q,\omega) = \frac{g_s}{L_y } \sum _{k_y} \frac {n _F
%(\epsilon_{n',s'} (k_y+q))-n_F (\epsilon_{n,s} (k_y) )} {
%\epsilon_{n',s'} (k_y+q) -\epsilon_{n,s} (k_y)-\hbar \omega}
%F_{ns,n's'}(k_y,k_y+q) \label{Pola}
\Pi _{n,n'} (q,\omega)   =   \frac{g_s}{L_y } \sum _{k_y} & & \frac
{f (\epsilon_{n'} (k_y+q))-f (\epsilon_{n} (k_y) )} { \epsilon_{n'}
(k_y+q) -\epsilon_{n} (k_y)-\hbar \omega} \times \nonumber \\
& & F_{n,n'}(k_y,k_y+q) \, \, \, ,\label{Pola}
\end{eqnarray}
where
$g_s$ is the spin degeneracy,  $f(\epsilon)$
is the Fermi distribution function,
$\epsilon_n(k_y)$ is the subband dispersion, and
$F_{n,n'}(k_y,k_y+q)$ is the square of the overlap between
wavefunctions,
\begin{equation}
F_{ns,n's'}(k_y,k_y+q)= \frac{1}{2} (1 + s s' \cos {\theta}),
\label{overlap}
\end{equation}
with  $\theta$ the angle between the wavevectors $(k_n,k_y)$ and
$(k_{n'},k_y+q)$. The Coulomb matrix elements for the armchair
nanoribbon take the form
\begin{eqnarray}
v  _{ijmn}(q)  = v  _{|i-j|,|m-n|}(qW)=\int _0 ^1 \! \! \! du \!
\int _0 ^1 \!  \! \! \! du' \nonumber \\    \cos [\pi(i \! - \! j)u]
\cos [\pi( m \!-\!n)u'] v(qW|u\!-\!u'|)\, , \label{Interactions}
\end{eqnarray}
where the one-dimensional Fourier transform of the the Coulomb
interaction has the form \cite{Li_1991}
\begin{equation}
v _q (x\! - \! x') =v (q |x\! - \! x'|)= \frac {2 e ^2 }{\epsilon
_0} K _0 ( q |x-x')|) \, \, .  \label{Coul_1D}
\end{equation}
Here $\epsilon _0$ is the background dielectric constant  and
$K_0(x)$ is the zeroth-order modified Bessel function of the second
kind, which diverges as -$\ln(x)$ when $x$ goes to zero.

In order to obtain the last expression in Eq. \ref{Interactions}
we have assumed that
the Coulomb interaction neither produces
intervalley scattering nor changes the sublattice index,
as should be appropriate in the long wavelength limit.  Note
that the interaction is strictly zero for arbitrary $q$ if
$|i-j|+|m-n|$ is an odd integer.

\begin{figure}
  \includegraphics[clip,width=8cm]{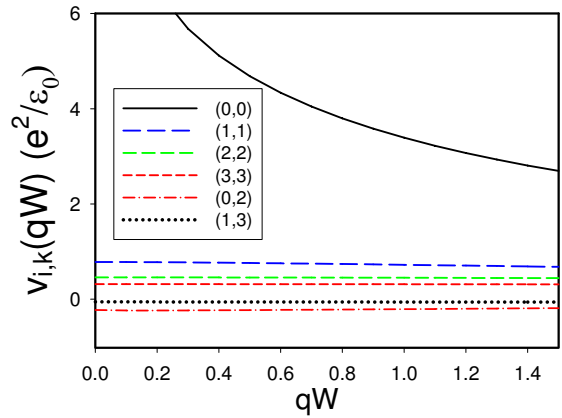}
  \caption{($Color$ $online$) Different matrix elements of the Coulomb interaction as function of
  $qW$ for a armchair terminated
  nanoribbon,  Equation \ref{Interactions}.}
   \label{Figure1}
\end{figure}

In Fig. \ref{Figure1} we plot different matrix elements of the
Coulomb interaction as a function of the momentum $q$. Clearly
$v_{0,0}$ is much larger than the others, and it diverges as $ -
\ln (qW)$ when $q \rightarrow $ 0. Thus, in graphene nanoribbons,
both in the undoped and doped cases, the
only relevant transitions at long wavelengths involve subband indices which satisfy
the condition $i-j$=$m-n$=0. For the low energy excitations in the
long wavelength limit we need only include the Coulomb matrix
element $v_{0,0}$, and can to a good approximation ignore the
others.

The charge density excitations are obtained by finding the zeros of
the real part of the determinant of the dielectric matrix
Eq.(\ref{diel}). These excitations are well defined provided they are
not damped by the continuum of electron-hole pair excitations,
defined by a finite imaginary part of the determinant of the
dielectric matrix.

In undoped graphene nanoribbons at zero temperature, only transitions from the
valence subbands, $s$=-1 to conduction subbands , $s$=1, contribute
to the polarizability.
Moreover, if we include
only $v_{0,0}$,
the lowest energy transitions are from the
valence ($s=-1$) to conduction ($s=1$) bands
which have the {\it same} subband index $n$.
The dielectric tensor has the form
\begin{eqnarray}
\varepsilon _{(n_1,s),(n_1,-s),(n_2,s'),(n_2,-s')} (q,\omega)=
\nonumber \\ \delta_{ss'}\delta_{n_1,n_2}- v_{0,0}(q) \Pi
_{(n_2,s'),(n_2,-s')} (q,\omega) \label{diel2}
\end{eqnarray}
The zeros of the determinant of this matrix are the
zeros of the function
\begin{equation}
1-v_{0,0} \sum _n (\Pi _{n-,n+}+\Pi _{n+,-n-}) \equiv 1-v_{0,0} \sum
_n \chi _n, \label{zeros}
\end{equation}
with
\begin{eqnarray}
\chi _n (q,\omega)= \frac{g_s}{L_y } \sum _{k_y} \frac {2 \Delta _n
(k_y,q) } { ( \hbar \omega+i\delta )^2 - (\Delta _n (k_y,q))^2}
\times \nonumber \\ F_{n-,n+}(k_y,k_y+q) \label{Polan}
\end{eqnarray}
%with $\hat {\epsilon} _{n} (k_y) = + \gamma a_0 \sqrt{k_n ^2 + k_y
%^2}$
where $\Delta _n (k_y,q)$=$|\epsilon _{n} (k_y+q)|$+$| \epsilon_{n}
(k_y)|$.

\begin{figure}
  \includegraphics[clip,width=8cm]{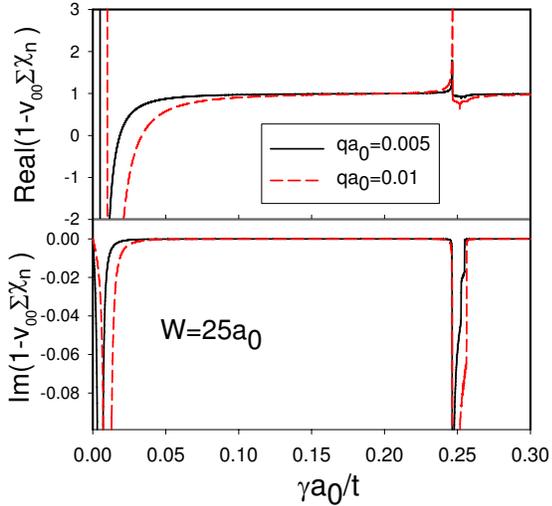}
  \caption{($Color$ $online$)Real and imaginary part of the dielectric constant,
  Eq. (\ref{zeros}),
   versus $\omega$
  for an undoped  metallic armchair nanoribbon of width $W$=25$a_0$.
  Dotted and dashed lines correspond to
  wavevectors
  $qa_0$=0.005 and $qa_0$=0.01 respectively.}
   \label{Figure2}
\end{figure}

In Fig. \ref{Figure2} we show the the real and imaginary parts of the
dielectric constant as function of frequency for an armchair
nanoribbon of width $W$=25$a_0$.
The lowest subband disperses linearly from zero
so that the nanoribbon is metallic.  Above this,
the next lowest energy subband occurs for
$k_1 \approx 0.13 a_0 ^{-1}$. In Fig.\ref{Figure2} we see that the
real part of the dielectric constant  is close to unity  and the imaginary
part has a non-vanishing value for frequencies larger than twice the
confinement energy of the first subband. The frequency ${\tilde
\omega} = \gamma a_0 ( \sqrt { k _1 ^2 + q ^2} + k_1)$ defines the
edge for the continuum of noninteracting intersubband electron-hole excitations.
Note that the dielectric constant does not cross zero
in the vicinity of these intersubband excitation energies.
The absence of such zeros indicates
there are no collective excitations associated with interband
transitions, which in any case for energies greater than $\hbar\tilde{\omega}$
would have a finite lifetime due to Landau damping.

For much lower
frequencies, the imaginary part of the dielectric function remains
zero down to an energy $ \sim \gamma a_0 q$, where a peak reveals the
existence of electron-hole excitations at a single frequency. This
corresponds to  an intraband excitation in the metallic subband.
(We note that in the lower panel of Fig. \ref{Figure2} this is slightly broadened due to the
small imaginary $i\delta$ added to $\omega$ to make the peak visible.
In the limit $\delta \rightarrow 0$ this peak becomes a delta function.)
%({\bf Luis: Did I get that right?})

\begin{figure}
  \includegraphics[clip,width=8cm]{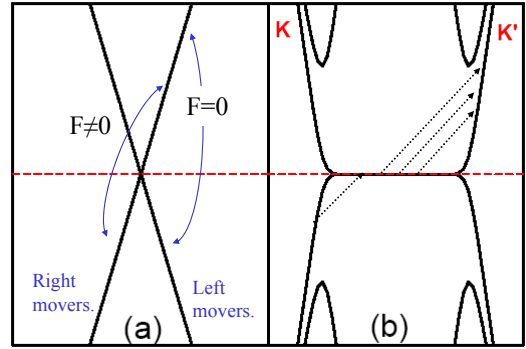}
  \caption{($Color$ $online$) Schematic representation of the band
  structure and low energy electron-hole
  pair excitations in intrinsic (a) metallic armchair nanoribbons and
  (b)zigzag nanoribbons.
  In metallic armchair nanoribbons the chirality of the
  wavefunctions allows only transitions between carriers
  moving in the same direction to enter the polarizability;
  contributions from backscattering vanish.
  For a given wavevector $q$,
  because of the linear dispersion,
  all the electron hole pairs entering the
  polarizability
  have the same energy $\gamma a_0 q$.
  In the case of zigzag nanoribbons the
  existence of surface states leads to
  a continuum in energies for electron-hole pairs with a given $q$.}
   \label{Figure3}
\end{figure}

The collapse of the particle-hole continuum into a single line
in the $\omega$ vs. $q$ plane is very common in one dimensional
metallic systems.  However, its presence here turns
out to depend in a fundamental way
on the chirality of the Dirac fermions.  This enters through the
overlap $F$ that appears in the polarizability, and guarantees
that for a given $q$ only particle-hole excitations of a single
frequency contribute to Eq. \ref{Polan}; {\it i.e.}, $\Delta_{n=0}$
is {\it independent} of $k_y$ when $F\ne 0$.  The overlap
$F=(1+\cos\theta)/2$, with $\theta$ the angle between
$(0,k_y)$ and $(0,k_y+q)$, is different than zero only
for $\theta=\pi$.
%({\bf Luis: I still think that a figure with band structure
%and arrows would be helpful.})
For excitations across the Dirac point, this corresponds to forward
scattering, with a single energy for a given $q$.  Although there is
a continuum of energies for a backscattered particle, these do not
contribute because the wavefunction overlap $F$ strictly vanishes.
This is illustrated in Fig. \ref{Figure3}(a).

{}From the top panel in Fig. \ref{Figure2} one may see that the real
part of the dielectric function vanishes at a frequency close to but
just above the particle-hole excitation frequency. Since the
imaginary part of the dielectric function vanishes at this
frequency, one obtains a single, sharply defined plasmon mode.
\emph{This is the only collective excitation that occurs in undoped
graphene nanoribbons.}

\begin{figure}
  \includegraphics[clip,width=8cm]{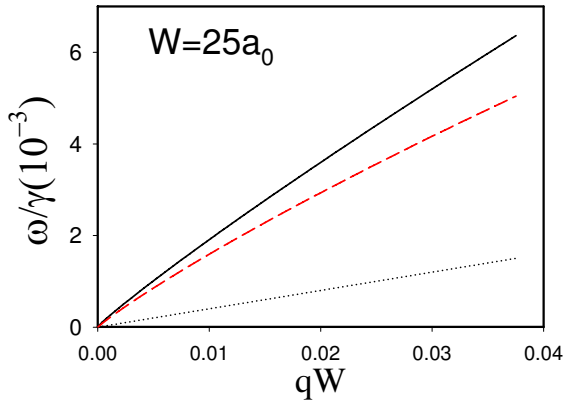}
  \caption{($Color$ $online$) The continuous line indicates the plasmon
  dispersion in a metallic graphene nanoribbon calculated within the RPA.
  The dashed  line shows the local longwavelength plasmon dispersion,
  Eq.\ref{plasmon}. The dotted line is the
  single particle excitation. }
   \label{Figure4}
\end{figure}

In Fig.\ref{Figure4} we plot the intraband plasmon frequency as a
function of $qW$.  It is clear in the figure that in the
long-wavelength limit, $qW<0.001$,  the Bessel function can be
approximated by -$\ln(qW)$ and the plasmon disperses  as $q
\sqrt{-\ln (qW)}$. Keeping only intraband transitions, the effective
polarizability entering in Eq. \ref{zeros} for  metallic armchair
nanoribbons can be obtained analytically for finite temperatures and
dopings,
\begin{equation}
\chi_0 (q,\omega,\beta,\mu)=  -\frac {g_s}{\pi} \frac{ \gamma a _0 q
}{(\hbar \omega ) ^2 -( \gamma a _0 q ) ^2}  \, f_1(q,\beta, \mu)
\end{equation}
with
\begin{equation}
f_1(q,\beta, \mu)=
 \frac {1} {\beta \gamma a_0} \left [ - \beta \gamma
a_0 q + 2  \ln { \frac {1 + e ^{-\beta \mu}}{ 1 + e ^{-\beta (
\gamma a_0 q + \mu)}}} \right ] \, \, ,
\end{equation}
with $\beta = 1/k_B T$ and $\mu$ the chemical potential.
This result applies both for doped and undoped  nanoribbons.
In the
longwavelength ($q \rightarrow 0$) limit we can easily show that the
plasmon frequency is given by,
\begin{equation}
\hbar \omega _p \simeq \left ( \frac {2 g_s e ^2}{\pi \epsilon _0}
\gamma a _0 q ^2 \right ) ^{1/2} \sqrt {- \ln (q W)} \sqrt
{f_1(q,\beta, \mu)} \, \, \, . \label{plasmon}
\end{equation}
Some comments on this result are in order. First, the plasmon has
exactly the same dispersion as the normal one-dimensional plasmon
of a metallic nanowire.
However, at zero temperature, in graphene nanoribbons the plasmon
frequency is independent of the chemical potential and therefore of
the density. A classical one dimensional plasmon has a dependence
$\sqrt{n_{1D}}$, where $n_{1D}$ is the one-dimensional carrier
density. A similar anomaly occurs in two dimensional doped
graphene where the plasmon frequency behaves as $(n_{2D})^{1/4}$,
compared  with    the $(n_{2D})^{1/2}$  classical
dependence \cite{Ando_2006,Wunsch_2006,Hwang_2006}. This anomaly in
the density dependence of the plasmons is a direct consequence of
the quantum relativistic nature of graphene \cite{Hwang_2006}.
Second, in undoped two dimensional graphene a finite temperature
produces thermo-plasmon excitations associated with transitions
between thermally activated electrons and
holes \cite{Vafek_2006,Apalkov_2006}. In metallic nanoribbons  the
absence of backscattering precludes the existence of such
thermo-plasmon excitations. The temperature only slightly alters the
plasmon dispersion. Finally, we stress the strong dependence
of the plasmon dispersion on the nanoribbon width. This
indicates that plasmon spectroscopy can be used as a
characterization of the width of the nanoribbon.

\subsection{Doped semiconductor armchair nanoribbon.}

\begin{figure}
  \includegraphics[clip,width=8cm]{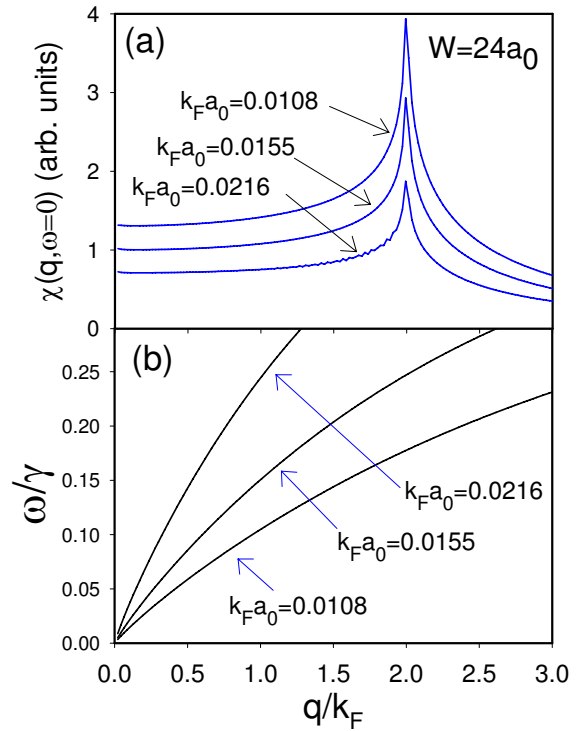}
  \caption{($Color$ $online$)(a) The static polarizability  as function of the wavevector of a doped semiconductor armchair nanoribbon for different
  densities. (b)Intrasubband plasmon dispersion of the same system.}
   \label{Figure5}
\end{figure}

We next analyze the low energy excitations of an electron-doped
armchair nanoribbon. We confine ourselves to the case that the Fermi
energy lies in the lowest energy subband. In this situation the
plasmon corresponds to excitations within the same subband. The
polarizability can be approximated by
\begin{equation}
\Pi _ {0,0} = \frac {2}{\pi} \int _{-k_F} ^{k _F}  \! \! \! \! \!d
k_y \, F_{0+,0+}(k_y,k_y+q)  \, \frac  {\Delta(k_y,q)} { (\hbar
\omega ) ^2 -  \Delta  (k_y,q)^2} \label{pi11}
\end{equation}
with $\Delta  (k_y,q)$=$|\epsilon _{1} (k_y+q)|$-$| \epsilon_{1}
(k_y)|$, and the collective exitations are obtained from the
equation
\begin{equation}
1-v_{0,0}  (q) \Pi_{0,0} (q,\omega) =0 \, \, \, \,
\end{equation}
The Fermi wavevector appearing in Eq.\ref{pi11} is related to the
one dimensional density of extra carriers in the nanoribbon by the
expression $k_F = \pi n_{1D}/2$. In Fig. \ref{Figure5}(b) we plot the
plasmon energy for this case and for different densities.  As
expected in one dimensional systems, the plasmon behaves as $q
\sqrt{-\ln {qW}}$. However, in contrast to the metallic case,
the low energy subband
dispersion is parabolic, and for small $q$ the overlap between
electron and hole wavefunctions, $F_{0+,0+}(k_y,k_y+q)$=$(1+\cos
\theta ) /2$, is close to unity.  Thus these plasmons have the usual
$\sqrt{n_{1D}}$ dependence on the electron density.  The possibility
of backscattering is also reflected in the static polarizability. In
Fig. \ref{Figure5}(a) we plot the static polarizability as a function
of the momentum. This quantity is logarithmically divergent at
$q$=$2k_F$ as in the standard one dimensional electron
gas. This is a consequence of the perfect nesting and the
possibility of backscattering from -$k_F$ to $k_F$.
It suggests that at sufficiently low temperature, the system
may be unstable to the formation of a charge density wave
when coupling to lattice phonons is included.

\subsection{Zigzag nanoribbon.}
The calculation of plasmon poles in the zigzag nanoribbon
case is complicated by the fact that the parameter
$z_n$ in the wavefunctions in Eq. \ref{zigzagWF}
is a function of $k_y$.  In this case the Coulomb
matrix element cannot be written only as a function
of the momentum transfer $q$.
The generalized  dielectric function
within the RPA obeys
\begin{eqnarray}
\epsilon_{ijmn} (k_y,k'_y;q,\omega) =
\delta_{im}\delta_{jn}\delta_{k_y,k'_y} -v_{ijmn}(k_y,k'_y;q)
\nonumber \times \\ g_s g_v  \frac {f (\epsilon_{m} (k_y+q))-f
(\epsilon_{n} (k_y) )} { \epsilon_{m} (k_y+q) -\epsilon_{n}
(k_y)-\hbar (\omega+i\delta)}\, \, \, . \label{diel1}
\end{eqnarray}
Here
$g_v$=2 is  the valley degeneracy.
The matrix element of the Coulomb interaction has the form
\begin{eqnarray}
v_{ijmn}(k_y,k'_y;q) \! = \! \int  _0  ^W  \! \! \!  dx   \! \!
\int_0 ^W
\! \! \! \!  dx' \Psi_{i,k_y} ^* (x) \cdot \Psi_{j,k_y+q} (x) \nonumber \times \\
v _q (x-x') \Psi _{m,k'_y+q} ^* (x') \cdot \Psi _{n,k'_y} (x')
\label{vmatrix1}
\end{eqnarray}
where $\Psi^* \cdot \Psi$ denotes a dot product of the vectors
appearing in Eq. \ref{zigzagWF}. The collective excitation spectrum
is obtained by the condition that the determinant of the dielectric
matrix, Eq. (\ref{diel1}), vanishes. In the case of undoped graphene
the dielectric function is different from unity only for transitions
from valence states, $s$=-1 to the conduction states $s$=1. Unlike
the armchair case, in
zigzag nanoribbons there is a broad particle-hole continuum of
excitations, from zero energy up to $\sim \gamma$, for small $q$.
This occurs because of the existence of surface states which are
essentially at zero energy over a broad range of $k_y$, allowing
arbitrarily low energy particle-hole excitations for any $q$
smaller than an appreciable fraction of the Brillouin
zone size, as illustrated in
Fig. \ref{Figure3}(b).
These
necessarily damp any plasmon poles that
might otherwise appear in the inverse dielectric function,
and we expect no propagating plasmon modes for
the undoped zigzag nanoribbon.

The doped case is essentially the same as that of the doped armchair nanoribbon
and standard semiconducting nanowires:
one may linearize the single particle spectrum around
the Fermi energy, leading to a narrow line of particle-hole
excitations, above which a plasmon pole should appear
with dispersion $q\sqrt{-\ln qW}$.
%{\bf (Luis: Should we say more?  I think the density
%dependence is more complicated than the doped armchair case
%because of the peak in the density of states.)}

\section{Summary}

In this paper we have analyzed the collective mode spectrum of
graphene nanoribbons.  In the undoped case, we find only metallic
armchair nanoribbons support a propagating plasmon mode.  This mode
mode may propagate undamped because the chirality of the wavefunctions
prevents it from decaying into particle-hole pairs.  We argued that
undoped zigzag nanoribbons will not support plasmon excitations
because this supports a broad continuum of particle-hole excitations
with low wavevectors, and there is nothing to protect a plasmon
from decaying into this.  Doped nanoribbons turn out to have properties
similar to those of semiconductor nanowires, including a plasmon
mode dispersing as $q\sqrt{-\ln qW}$ with a coefficient that vanishes
when the doping vanishes (except in the metallic armchair case), and
a static dielectric response that is divergent at $q=2k_F$.

\vspace{0.5truecm}

{\bf Acknowledgements.}  The authors thanks P. de Andr\'es, F.Guinea
and P.L\'opez-Sancho  for helpful discussions. This work was
supported by MAT2006-03741 (Spain) (LB) and by the NSF through Grant
No. DMR-0454699 (HAF).

%\bibliography{mia}

\end{document}